# SLOW DYNAMICS IN AN AZOPOLYMER MOLECULAR LAYER STUDIED BY X-RAY PHOTON CORRELATION SPECTROSCOPY

**Davide Orsi[1], Luigi Cristofolini[1], Marco P. Fontana[1], Anders Madsen[2], Andrei Fluerasu[3]**

[1]Physics Department, University of Parma, Viale Usberti 7/A, Parma 43100, Italy
[2]European Synchrotron Radiation Facility, 6 rue J. Horowitz, 38043 Grenoble, France
[3]Brookhaven National Laboratory, NSLS-II, Upton NY 11973, USA

**ABSTRACT**

We report the results of X-ray photon correlation spectroscopy (XPCS) experiments on Langmuir Blodgett multilayers of a photosensitive azo-polymer. Time correlation functions have been measured at different temperatures and momentum transfers (q) and under different illumination conditions (darkness, UV or visible). The correlation functions are well described by the Kohlrausch-Williams-Watts (KWW) form with relaxation times that are proportional to $q^{-1}$, which in other systems have been explained in terms of intermittent rearrangements [L. Cipelletti *et al.*, Phys. Rev. Lett. **84**, 2275-2278 (2000)] or random dipolar interactions within an elastic medium [J.-P. Bouchaud and E. Pitard, Eur. Phys. J. E **6**, 231-236 (2001)]. The characteristic relaxation times follow the well known Vogel-Fulcher-Tammann law describing the temperature dependence of the bulk viscosity of this polymer. UV photoperturbation accelerates the relaxation dynamics, in qualitative agreement with the fluidification effect of UV photo-perturbation previously observed by surface rheometry, and is used to drive the system out of equilibrium. Transient dynamics is characterized, by the variance $\chi$ of the two-times correlation functions. A clear peak in $\chi$ appears at a well defined time $\tau_C$ which scales with $q^{-1}$ and with the ageing time, in a similar fashion as previously reported in colloidal suspensions [O. Dauchot *et al.*, Phys. Rev. Lett. **95**, 265701 (2005)]. From an accurate analysis of the correlation functions we could demonstrate a temperature dependent cross-over from KWW compressed to simple exponential behaviour, which is modified by the fluidification due to the optical pumping of the cis-trans isomerisation of the side-chain azobenzene group.

## I. INTRODUCTION

The changes in dynamical properties of azobenzene containing polymeric layers following optical pumping of the trans-cis isomerization transition have received much attention in past years, mainly due to promising applications in optical writing with the potential to achieve nanometre resolution [1]. However -up to now- end user applications for nanowriting have not been devised, mainly because of an insufficient understanding of the fundamental properties of these intriguing, though complex, systems. They feature a rich phenomenology which is also of more general relevance, *e.g.* to phase transitions in liquid crystals, the glass transition, de-wetting phenomena, and anomalous diffusive and vibrational dynamics in spatially heterogeneous systems. For example, non-adiabatic dynamical simulations of photoinduced processes in a photosensitive liquid crystal have shown remarkable differences in the photoswitching process of azobenzene in the bulk with respect to the same dynamics of the isolated molecule. This evidences the influence of the condensed phase environment on the mechanisms governing the time scale of photoswitching [2].

All of these issues can be studied in a specific class of materials, namely side-chain liquid crystalline polymers with an azobenzene moiety attached to the main chain via a flexible molecular spacer [3-5]. In poly-acrylates and poly-methacrylates the pumping of the photo-isomerization transition yields several potentially useful optical effects with very high sensitivity [6-8]. The vibrational and diffusive dynamics can be altered by the optical pump [9,10] and the phenomenology of the glass transition has been extensively studied by calorimetric, dynamical analysis and ESR relaxation measurements [11]. Extensive photo-mechanical effects have also been observed in azopolymers, including photo induced expansion and a change in the viscoelastic properties [12]. The use of photo-isomerization to produce actuator materials was first proposed by de Gennes [13] and since then a wide literature flourished; however, very few real applications emerged. At present, photoinduced effects on the visco-elasticity of polymers are poorly understood but in the future they could play a key role for a number of new applications in different fields, including specialized fabrics and drug delivery. A recent work by Ketner *et al.* [14] revealed the remarkable property of aqueous micellar solutions, the viscosity of which can be reduced by more than four orders of magnitude upon UV exposure. The limiting factor in that study was the irreversible character of the transformation; however, in a very recent study [15] reversible rheological properties of an azo-containing siloxane polymer backbone were demonstrated. Such properties are of particular interest for light-activated damping mechanisms, actuable armour, and related applications in fields such as robotics and sensors.

Previously, the Langmuir-Blodgett (LB) technique [16] was applied to obtain monolayers and multilayers of poly[[4-pentiloxy-3'-methyl-4'-(6-acryloxyexyloxy)] azobenzene] henceforth called PA4, which has a bulk glassy phase with $T_g = 20°C$ and a nematic phase with clearing point $T_{NI}=92°C$ [17]. This allowed studying the effects of thickness on the structure and dynamics in samples down to the nanoscale [18]. This is of key interest, not only to explore the potential of nanowriting applications, but also to scrutinize a fundamental problem of long standing in the dynamics of disordered systems, namely the behaviour of glasses in confined geometries or with reduced dimensionality. Here, one of the central assumptions is the existence of so-called cooperatively rearranging regions [19,20] and a model providing the quantitative foundation of the classical Adam-Gibbs ansatz [21]. As the size of the system crosses the characteristic length scale of such regions, a disruption of the collective Adam-Gibbs behaviour is expected. Several experimental studies have been performed either in glassy systems confined within nanopores [22] or in ultrathin films [23-26]. The effects of restricted dimensionality on the glass transition of PA4 multilayers as a function of temperature were previously investigated [18]. The samples were perturbed by pumping the trans-cis photo-isomerization transition and the relaxation process back to equilibrium was followed using real time null-ellipsometry on samples of different thicknesses, ranging from one to sixty monolayers. A crossover in the dynamical behaviour was found for samples with thicknesses corresponding to approximately eight monolayers. This was interpreted as being caused by the formation of a "dead layer" close to the substrate with a much slower dynamics than the bulk of the polymer. In a following experiment [27] the effect of temperature and photo-perturbation were investigated on thick films and on superstructures formed by the intercalation of fatty acids (Ba Behenate) and azopolymer molecular layers. This work demonstrated the possibility of generating nanostructured glasses with a controlled size and morphology.

In the present work we take advantage of this possibility to study slow dynamics inside a photosensitive azopolymer as a function of temperature and photoperturbation, exploiting the spatial resolution in X-ray photon correlation spectroscopy (XPCS) experiments [28]. The main advantage of using X-rays instead of direct methods (*e.g.* scanning probe microscopy) is that it provides statistical information averaged over the whole sample as a function of the momentum transfer q. This is essential for the analysis of dynamical heterogeneity and of non-equilibrium and aging effects in the observed dynamics. Indeed, heterogeneities have been observed in recent molecular dynamics (MD) simulations [29] with a pronounced effect of photo-isomerization on the non-equilibrium dynamics of glassy and supercooled molecular matter. In the MD studies the effect of isomerization is to create mobile regions of dynamical heterogeneity inside the matrix hence leading to an important increase of the diffusion coefficient and a liquid-like behaviour below the glass-transition temperature.

In general, systems out of equilibrium often display very complex dynamics that is far from being described by simple Brownian motion. Examples include clay suspensions [30,31], colloidal gels [32-35] or concentrated emulsions [36]. In all these systems, the bulk dynamic structure factor is well described by a peculiar universal form – compressed exponentials – and shows a complex time dependence (i.e. "aging").

In recent papers [37, 38] XPCS was used to measure the dynamic structure factor of gold nanoparticles moving on the surface of thin polymer films. Above the glass transition of the polymer a stretched or compressed KWW behaviour of the correlation functions was observed depending on the thermal history of the sample. The relaxation rates appeared to scale linearly with q, excluding a simple Brownian diffusive motion. Excluding over-damped capillary waves as the origin of the linear behaviour, the authors drew resemblance with aging bulk soft matter systems governed by a power law distribution of particle velocities due to ballistic motion. However, since the sample was prepared by evaporation of gold on the already deposited polystyrene film, it cannot be ruled out that the film is driven out of equilibrium by the way it was prepared. Therefore further investigations are required to clarify the role of aging.

## II. EXPERIMENT

The XPCS experiments were performed using partially coherent X-rays at the ID10A beamline (Troika) of the European Synchrotron Radiation Facility. A single bounce Si(111) crystal monochromator was used to select 8.06 keV X-rays, having a relative bandwidth $\Delta\lambda/\lambda \approx 10^{-4}$. A Si mirror downstream of the monochromator was used to suppress higher energy X-rays from the monochromator. A transversely coherent beam was defined by slit blades with highly polished cylindrical edges. The slit size was 10 x 10 $\mu m^2$. A set of guard slits was placed just upstream of the sample to block the parasitic scattering due to diffraction from the beam-defining slit. The scattering from the sample was recorded by the Maxipix detector [39] which consists of 256x256 pixels each of 55x55 $\mu m^2$ area, located 2.2 meters downstream of the sample. In this way q ranging from 0.005 to 0.2 $nm^{-1}$ was accessible. Focusing on the slow dynamics, we collected speckle patterns every 1-10 sec, being limited on the long time side only by the overall instrumental stability yielding an upper limit of a few hours in the accessible timescales.

Based on previous experience we employed a particular experimental geometry optimizing both the efficiency of the photoperturbation and the XPCS data collection as explained in the following. The polymer film was deposited as a 75 molecular layer thick Langmuir-Blodgett multilayer on a silicon substrate. It was extensively characterized by null-ellipsometry, AFM and SEM microscopy, evidencing a typical roughness of 1 nm and a thickness of 150 nm, in excellent agreement with the typical values previously reported for LS films of this polymer [18, 27]. The sample was oriented to have the X-ray beam impinging at a grazing angle of 0.15°, *i.e.* well below the critical angle $\alpha_c$ for total external reflection from the silicon-air interface at 8.06 keV ($\alpha_c$=0.22°). This allows measurements of small angle scattering from the polymer around the specular reflection, and at the same time it minimizes radiation damage to the polymer sample caused by X-ray induced photoelectrons emitted by the Si substrate. The correlation functions were averaged over the out-of-plane direction, i.e. perpendicular to the scattering plane, thus focusing on the in-plane signal to probe the dynamics inside the film and exclude any contribution from corrugation of the surface. We focused on a few selected temperatures well above $T_g$, between 42° and 85°C. Attempts to investigate higher temperatures proved unsuccessful because the polymeric molecular multilayer starts dewetting from the substrate and forms droplets. We explored different illumination conditions: dark, UV ($\lambda$ ~360 nm) and visible light ($\lambda$ ~480 nm). UV illumination is expected to alter the cis-trans equilibrium in favour of the cis configuration, whereas at 480 nm both cis and trans configurations are pumped with roughly the same efficiency, yielding a continuous cycling of the azobenzene molecules between the two states.

For the data analysis we proceeded as follows: first, individual pixels of the Maxipix detector were grouped in 15-25 q-bins according to the values of their in-plane momentum transfer q. Subsequently, every q-bin was filtered by excluding those pixels with average signals or standard deviations significantly different from the average values. The intensity auto-correlation functions $g^{(2)}(q, t)$ defined as

$$g^{(2)}(q,t) = \frac{\langle I(q,t_0) * I(q,t_0+t) \rangle}{\langle I(q) \rangle^2}$$

were then calculated. Here $I(q,t_0)$ is the intensity measured at time $t_0$ and wave vector q and the average <...> is performed over all the pixels in the q-bin and eventually over all initial times $t_0$ (for equilibrium dynamics).

The correlation functions were then fitted by a Kohlrausch-Williams-Watts (KWW) modified exponential form

$$g^{(2)}(q,t) = A(q) + \beta(q) * e^{-2(t/\tau)^\gamma}$$

where $A(q)$ is the baseline, typically equal to unity, $\beta(q)$ is the contrast of the correlation function which in this geometry is around 0.3. $\tau$ is the characteristic relaxation time that usually follows a power law as a function of q with exponent n: $\tau \propto q^n$ and $\gamma$ is the KWW exponent.

### III. RESULTS AND DISCUSSION

**Equilibrium dynamics:** We first focus on the dynamics in the two different equilibrium configurations of the material either in dark or under UV illumination limiting the investigations to the long time stationary regimes. The out-of-equilibrium dynamics shall be discussed in the next section.

The first effect of UV photoperturbation is to alter the cis/trans isomer ratio. This happens typically on a sub second timescale [6]. A slower mechanical expansion of the film takes place, on a time scales ranging up to 100 s at 60°C [40]. All the results reported in this section were obtained long time after the application of photoperturbation. Typical correlation functions, together with their best fits are shown in Figure 1 for PA4 at 72°C after it has reached its equilibrium state under UV illumination.

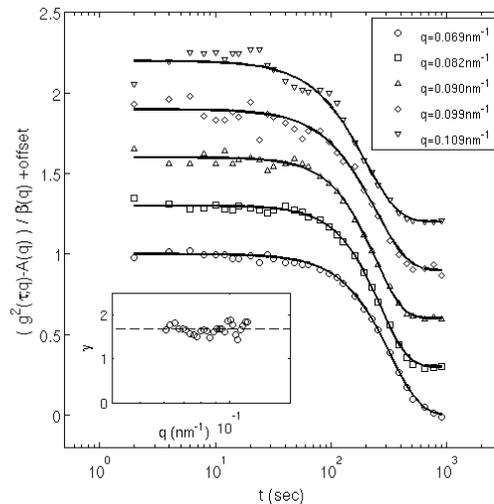

**Figure 1** *Normalized correlation functions for PA4 at 72°C under UV illumination, together with the best KWW fits (line). Data are offset in steps of 0.3 for clarity. **inset**: best fitting values for the compression exponent $\gamma$.*

Each correlation function was fitted with the KWW form with all parameters left free to vary. In the subsequent analysis we heuristically rejected about 10% of the data corresponding to the worst fits as measured by the $\chi^2$ test. The extracted relaxation times $\tau$ are shown in figure 2 as a function of q for different temperatures and illumination conditions.

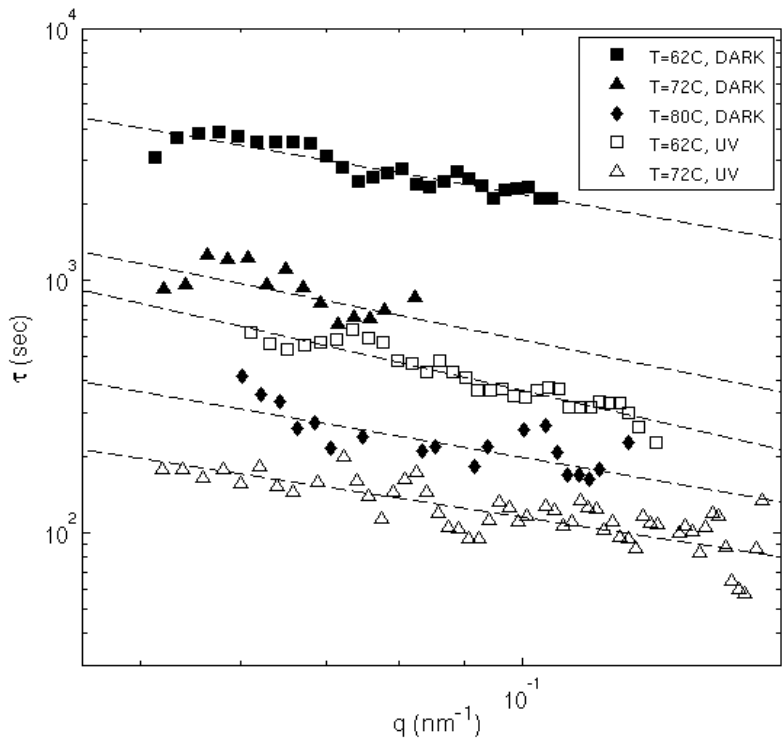

*Figure 2*: *Equilibrium relaxation times $\tau$ as a function of q at the different temperatures and illumination conditions indicated in the caption.*

Power law fits are shown by the dashed lines in figure 2 which clearly shows a speeding-up of the dynamics upon UV illumination. As shown in figure 3 (top left) the exponent n is on the average -1 for all temperatures both in dark and under UV illumination. The KWW exponent $\gamma$ shows no significant q dependence (see inset of figure 1 for an example) but it depends on the temperature, with a correlation function decaying faster than exponentially ($\gamma$=1.5, compressed exponential behaviour) at the lower temperatures and a value closer to $\gamma$=1 (simple exponential) at the highest temperatures (see figure 3, top right panel). At the same time the contrast $\beta$ and the baseline A(q) varied slightly from one correlation function to another without any apparent regularity but with $\beta$ always close to 0.3 and A(q) around 1, as expected.

A similar $\tau \sim q^{-1}$ scaling and compressed exponentials, i.e. $\gamma$>1, has been measured by XPCS in a wide variety of materials [41], including both attractive (fractal colloidal gels [32]) and repulsive (compact arrangements of soft elastic spheres [42], of emulsion droplets [43], and Laponite [30]) systems. This scaling suggests that the slow dynamics is due to a ballistic motion in the samples, in the sense that the average mean-square displacement grows linearly with time. It has been proposed [32, 42, 43] that the dynamics could be due to randomly distributed stress sources within the sample, whose response mimics that of an elastic solid. The dynamics may therefore be caused by a series of discrete rearrangement events as discussed in a number of publications [41]. The mechanism underlying this is at present unclear: for colloids and related systems, following Cipelletti, one can postulate intermittent rearrangement of larger or smaller volumes whose size can be estimated by some "crossover" q at which the compression coefficient varies from 1 to 1.5 [41]; however the situation is much less clear for polymeric systems. Here predictions have been formulated based on a model by Bouchaud and Pitard concerning an elastic medium with random dipolar interactions [44] with a crossover from $\gamma=3/2$ to $5/4$ at some characteristic time scale $\tau_q$. We assume $\tau_q$ to be temperature dependent and to decrease upon heating following the temperature dependence of the viscosity. We can then interpret our experimental result of the decrease of $\gamma$ upon heating (above 80 °C for *trans* PA4, dark, and above 65 °C for *cis* PA4, UV) as support for such a crossover.

Moreover, the different crossover temperatures for dark and UV illuminated PA4 (approximately 15 °C difference), clearly show that UV photo-perturbation accelerates the molecular dynamics, as demonstrated by the data in the bottom left panel of figure 3. This photoinduced acceleration can be quantified by the ratio of the relaxation times under UV illumination and in dark ($\tau_{UV}/\tau_{DK}$), which is, within our experimental accuracy, independent of T and on the order of 0.2. This is in qualitative agreement with Interfacial Shear Rheometry (ISR) measurements on a PA4 Langmuir film in the same illumination conditions, where we found a reduction of the dynamical shear modulus $|G^*|$ by a factor of ~7, as shown in figure 3, bottom right panel. However, a quantitative comparison of the results is difficult because the relation between $|G^*|$ and the non-diffusive dynamics observed by XPCS is not trivial [45].

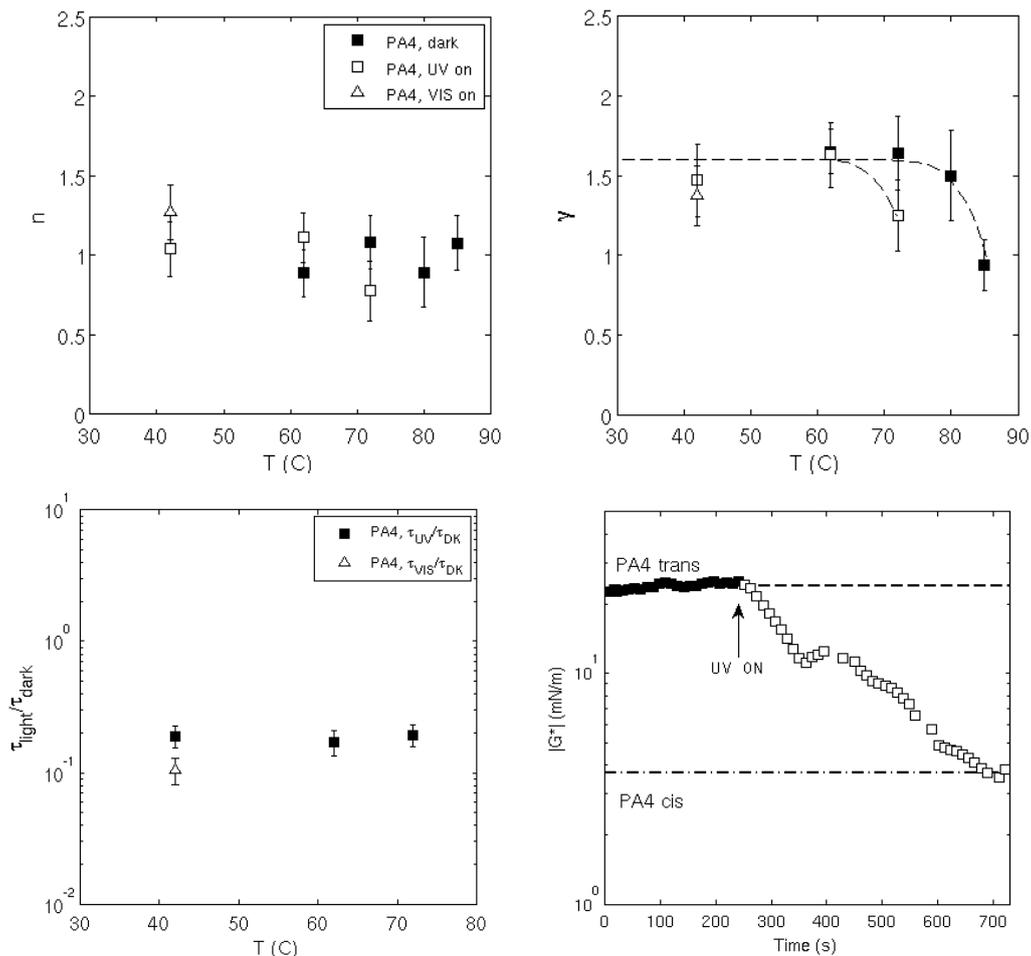

**Figure 3: top left** *Temperature evolution of the exponent n at different illumination conditions;* **top right** *same for $\gamma$ averaged over the investigated q range; the dashed lines are guides to the eye.* **bottom left:** *XPCS measurement of the photoinduced acceleration of dynamics, quantified by the ratio of relaxation times ($\tau_{UV}/\tau_{DK}$, $\tau_{vis}/\tau_{DK}$)* **bottom right**: *Interfacial Shear Rheometry measurement of photoinduced fluidification: evolution of the shear modulus $|G^*|$ in a Langmuir molecular layer of PA4 in dark and under UV light.*

We also performed measurements under visible light which, from ISR data, is known to increase the polymer's shear modulus, at least on the macroscopic scale. Surprisingly we found a *decrease* of the relaxation time comparable or even larger than that obtained by UV illumination (Fig. 3). We speculate that this may be explained by the different length sensitivity in XPCS and ISR measurements. In XPCS $2\pi/q$ sets the length scale which is smaller than ~150 nm in all the measurements presented here. Hence the local scale dynamics which is probed here apparently becomes faster due to the isomerisation cycles induced by the visible light illumination (both *cis to trans* and *trans* to *cis* isomerisation processes have approximately the same efficiency at this wavelength). At the same time by ISR we observe this cyclic isomerisation increasing the shear modulus of the film. ISR is sensitive to macroscopic length scales (mm) and hence we speculate that in ISR, in contrast to the XPCS result, the film stiffens by the cis-to-trans cycles because they reduce defects and grain boundaries, similarly what happens during an annealing process.

Finally we analyzed the temperature dependence of the relaxation times, as shown in the Arrhenius plot of figure 4. We compare our data with the Vogel-Fulcher-Tammann (VFT) law

$$\tau = \tau_\infty \exp\left(\frac{T_A}{T - T_0}\right)$$

describing the viscosity of PA4 ($T_0 = -30$ °C, activation temperature $T_A = 997$ °C) [46] which was confirmed by EPR, depolarized micro Raman and Quartz Crystal Microbalance experiments [9]. Inspection of Fig 4 shows that the same VFT law is obeyed by XPCS relaxation data both in dark conditions and under UV illumination, excluding the points recorded at the lowest studied temperature (42 °C), where the slow dynamics may start approaching the long-time stability of the experimental setup because of phenomena such as the drift of the synchrotron X-ray beam (for the dark data) or radiation induced damage of the sample for the UV case.

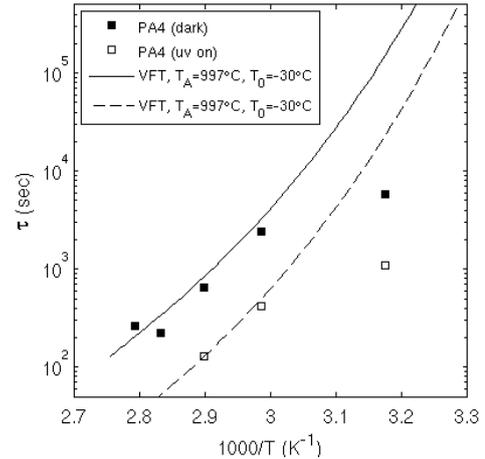

**Figure 4:** *Arrhenius plot of tau in dark, UV illumination, all at the same value of q=0.09 nm$^{-1}$ together with the well characterized VFT law for PA4.*

**Out of equilibrium dynamics:** In this section the photosensitivity of molecular films of PA4 is exploited to drive the system out of equilibrium and to study the dynamics in this state. To characterize the out of equilibrium dynamics, as pioneered by Sutton and co-workers [35, 47, 48], the usual time averaged correlation function has to be substituted by a two-times correlation function $g(q,t_0,t_1)$ defined as:

$$g(q,t_0,t_1) = \frac{\langle I(q,t_0)I(q,t_1)\rangle}{\langle I(q,t_0)\rangle\langle I(q,t_1)\rangle}.$$

In figure 5 we report $g(q,t_0,t_1)$ in a typical out of equilibrium situation when the photoperturbation is switched off. Larger values of correlation are represented by darker shades, as displayed in the greyscale. In this representation, the ageing time $t_{age}$ is given by $(t_0+t_1)/2$ (i.e. proportional to the displacement along the diagonal indicated by the arrow in the figure) while the lag time t is given by the absolute difference $|t_1-t_0|$ (i.e. the distance of a point from the diagonal). Stationary equilibrium dynamics would correspond to contour lines running parallel to the diagonal. On the contrary, as it is the case in the figure, contour lines moving away from the diagonal vs. age indicate a slowing down of the dynamics. This is caused by switching off the UV photo-perturbation which leaves the system to relax back towards its equilibrium trans configuration where the relaxation is much slower, see Fig. 3.

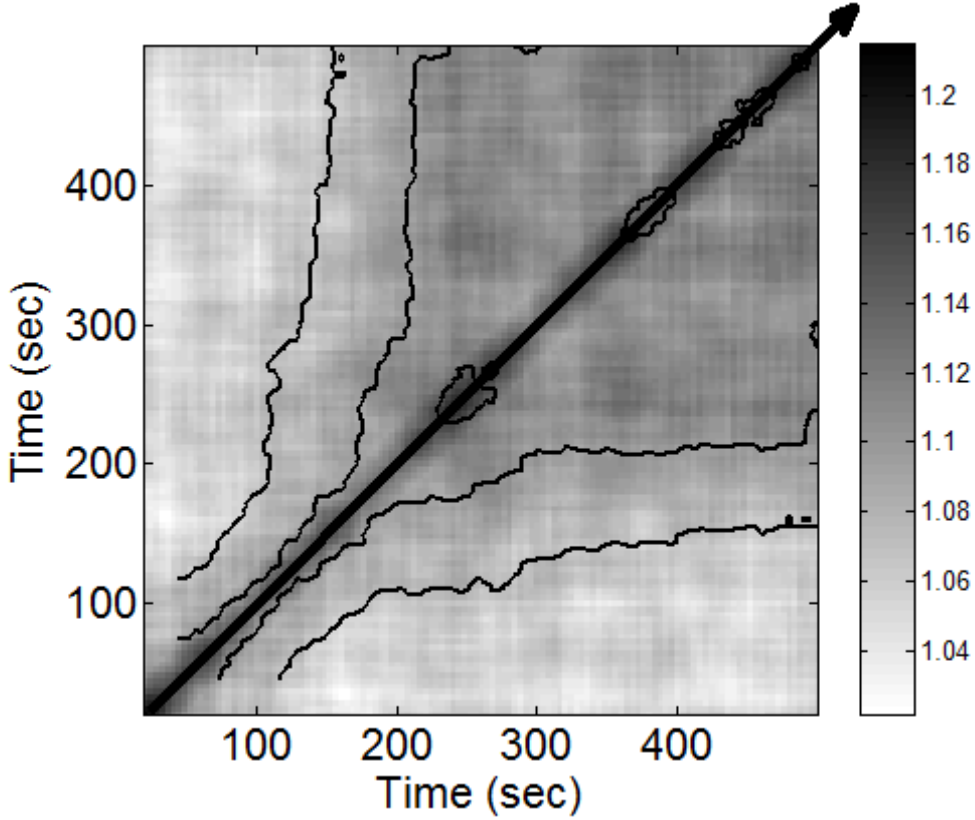

**Figure 5:** *Two-times correlation function during the cis-to-trans transformation at 72 °C. UV light was switched off at the initial time t=0. The arrow indicates the diagonal along which ageing time increases.*

A useful measure of the fluctuations of two-times correlation functions is provided by the normalized variance $\chi(q, t)$ defined as in ref [34]:

$$\chi(q,t) = \frac{\langle g^2(q,t_0,t_0+t)\rangle_{t_0} - \langle g(q,t_0,t_0+t)\rangle_{t_0}^2}{(g^{(2)}(q,t=0)-1)^2}$$

where the average <...> is performed on all the initial times $t_0$ while the zero time correlation $g^{(2)}(q, t=0)$ provides the appropriate normalization.

This quantity depends obviously on q and on the delay time t, and is in many ways similar to the dynamical four-point susceptibility $\chi^{(4)}$ that has been widely studied in glassy materials to characterize spatial heterogeneity in their dynamics [49, 50]: while the quantity $\chi$ measured by XPCS characterizes temporal heterogeneity and $\chi^{(4)}$ characterizes spatial heterogeneity, the two are related noting that an increasing length scale over which the dynamics is cooperative would imply fewer dynamically independent scattering sites in the given scattering volume and therefore an increase in temporal fluctuations measured by $\chi$ [35]. Specifically, in the case of homogeneous diffusion dynamics, $\chi$ is independent on the delay time, while an increase in $\chi$ has been identified as a signature of growing dynamical correlation lengths in a range of disordered systems approaching arrest, such as supercooled liquids near the glass transition and granular materials near the jamming transition.

However when the system is off equilibrium several factors might contribute to $\chi$. In particular in the case shown in figure 5, the slow evolution of the correlation function as a function of the ageing time can be understood in terms of the general slowing down of the dynamics of the system after switching off the photoperturbation. This slow dependence is fitted by a polynomial expression and subtracted before any further analysis.

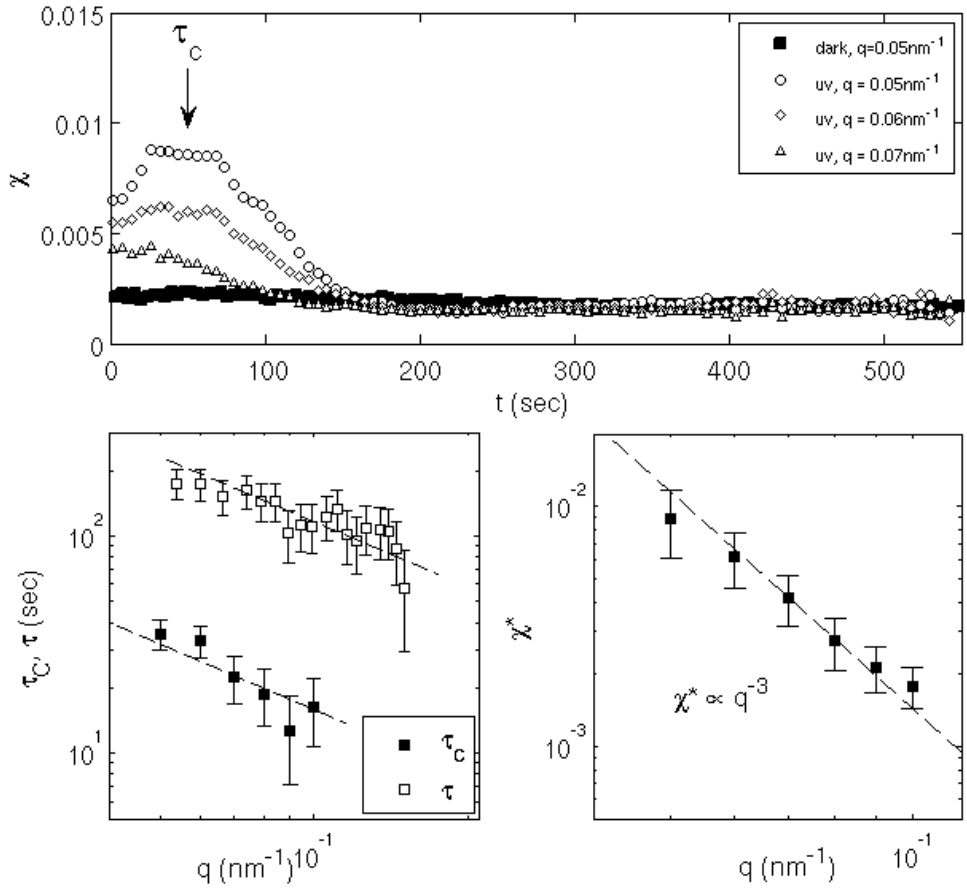

**Figure 6 top:** *variance of the two times correlation function for PA4 at 72 °C under UV illumination and in dark. Note the presence of a peak only for the UV data.* **bottom left:** *q dependence of the peak time $\tau_c$ for the UV data compared to that of the relaxation time $\tau$. Dashed lines indicate the $\tau \propto q^{-1}$ power laws discussed in the text* **bottom right:** *q dependence of the amplitude of the peak $\chi^*$.*

In figure 6, top panel, we report the evolution of χ for different values of q, as a function of lag time t for PA4 at 72 °C in dark and under UV illumination. A well defined peak in χ centred at time $\tau_c$ and with amplitude $\chi^*=\chi(\tau_c)$ emerges when the sample is exposed to UV light. No similar peak was observed in dark at any q or at any temperature. Focusing on the UV data, the characteristic time $\tau_c$ follows a similar scaling relation $\tau_c \approx q^{-1}$ as the relaxation time τ (bottom left panel). The ratio $\tau/\tau_c$ is about 5 like in previous experiments [33] on ageing colloidal systems. We also followed the q-dependence of the intensity of the peak $\chi^*$ (bottom left panel), which is expected to be inversely proportional to the number of independent regions in the scattering volume that are responsible for the observed dynamical heterogeneity. We find $\chi^* \propto q^{-3}$, which is reasonable assuming the size of the regions to be proportional to the space scale probed (d=2π/q) at each momentum transfer q.

In figure 7 we summarize the results of our analysis of the dynamics in the transient phase induced by the UV photoperturbation, which was switched on at time $t_{age}=0$. In the top panel we show the variance χ for different ageing times $t_{age}$: the peak characteristic time $\tau_c$ –shown in the bottom left panel- increases at early age, reaching a stationary value after about 200 seconds while the peak amplitude $\chi^*$ -bottom right panel- remains roughly constant at all ages. This is rationalized as follows: as anticipated photomechanical effects lasting about 100-200 seconds [40] dominate relaxation at early times, prevailing the pure diffusional motion.

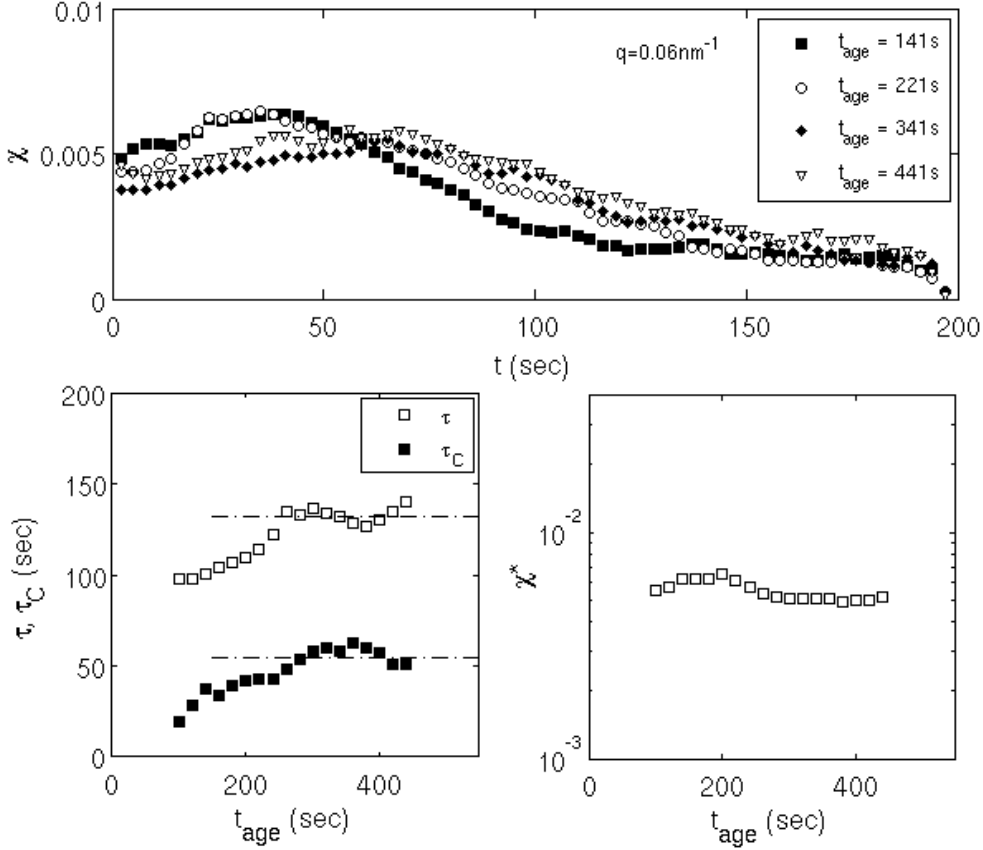

**Figure 7 top:** *Off equilibrium evolution of the variance χ for PA4 at 72 °C. UV photoperturbation was switched on at $t_{age}=0$. Note the shift of the peak as a function of the ageing time* **bottom left:** *ageing dependence of the peak time $\tau_c$ compared to that of the relaxation time $\tau$. Dashed lines indicate their corresponding equilibrium values* **bottom right:** *amplitude of the peak $\chi^*$ at different ageing times*.

At longer times these transient effects disappear and both the relaxation time and the characteristic time $\tau_c$ reach their equilibrium values, indicated by the dashed horizontal lines in the bottom left panel of the figure. We emphasize the strict correspondence found in the behaviour of these two in principle independent quantities. This is in some sense analogous to what observed in a different system, namely in colloidal gels recovering from shear: also in that case both times $\tau$ and $\tau_c$ follow the same law, which in that case is a linear increase with ageing time [35, 51].

Concerning the measurements in dark, the absence of a peak in $\chi$ can be explained assuming that the dynamics in this case is homogeneous at least on the length scale probed in the experiment. This could arise e.g. because the heterogeneity develops on a much slower timescale than that accessible in our experiment, or could be an effect related to the presence of the nematic potential which reduces the dynamical heterogeneity intrinsic of the supercooled glassy phase. The implications of this observation are currently under investigation.

## IV. CONCLUSIONS

In this paper we have shown an unconventional application of XPCS that allows to determine local dynamical changes caused by the cis-trans isomerisation in an azopolymer Langmuir Schaefer thin film, and we relate the findings to current theories about non-equilibrium processes in glassy systems. In particular we were able to determine the scaling laws for equilibrium and non-equilibrium fluctuations on local (nanometric) space scales in the long time limit (up to several hours). The data are in qualitative agreement with the random stress model developed by Bouchaud et al. [44] and we could bridge the gap between local and macroscopic scales by showing a qualitative agreement in the behaviour of the XPCS relaxation times with the light induced changes in the macroscopic shear modulus (on a shorter time scale) obtained by ISR. Moreover, the temperature dependence of the relaxation times follows the same Vogel-Fulcher-Tammann law describing the viscosity of the polymer.

An accurate insight into the non-equilibrium dynamic of the system comes from the analysis of the variance of the two times correlation functions. For *cis*-PA4 it presents a peak at a characteristic time $\tau_c$, with an inverse proportionality between $\tau_c$ and q similar to that of the relaxation time $\tau$. These features are commonly accepted in the literature as hallmarks of heterogeneous dynamics, and have been observed previously in a few model systems. At the same time, in *trans*-PA4, no peak was observed. This is possibly due to a suppression of the dynamical heterogeneities by the presence of a nematic potential.

## ACKNOWLEDGEMENTS

We acknowledge C. Caronna, E. Pontecorvo and F. Zontone for help with the experiments.